\documentclass[10pt, twocolumn]{IEEEtran}
 \usepackage{amsmath,amssymb}
 \usepackage{subfigure}
 \usepackage{graphicx,graphics,color,psfrag}
 \usepackage{cite,balance}
 \usepackage{caption}
 \allowdisplaybreaks
 \usepackage{algorithm}
 \usepackage{accents}
 \usepackage{amsthm}
 \usepackage{bm}
 \usepackage{algorithmic}
 \usepackage[english]{babel}
 \usepackage{multirow}
 \usepackage{enumerate}
 \usepackage{cases}
 \usepackage{stfloats}
 \usepackage{dsfont}
 \usepackage{color,soul}
 \usepackage{amsfonts}
 \usepackage{cite,graphicx,amsmath,amssymb}
 \usepackage{subfigure}
   \usepackage{booktabs}
 \usepackage{fancyhdr}
 \usepackage{hhline}
 \usepackage{graphicx,graphics}
 \usepackage{array,color}

 \usepackage{amsmath}

 \usepackage{lipsum}
\usepackage{mwe}


\newcounter{parentnumber}

\def\phi{\varphi}

\def\({\left(}
\def\){\right)}

\def\b0{{\mathbf{0}}}

\begin{document}

\title{\Huge V2X-Based Vehicular Positioning: \\  Opportunities,  Challenges, and Future Directions}

\author{Seung-Woo Ko,  Hyukjin Chae,   Kaifeng Han,  Seungmin Lee,  Dong-Wook Seo, and Kaibin Huang  \\
\thanks{S.-W. Ko and D.-W. Seo are with Korea Maritime and Ocean University, Busan, Korea (\{swko, dwseo\}@kmou.ac.kr). H. Chae is with Ofinno LLC, VA, USA (hchae@ofinno.com).
K. Han is with China Academy of Information and Communication Technology, Beijing, China  (hankaifeng@caict.ac.cn). S. Lee is with LG Electronics, Seoul, Korea (edison.lee@lge.com).  K. Huang is with The University of Hong Kong, Hong Kong (huangkb@eee.hku.hk). }
}
\maketitle

\begin{abstract} 
\emph{Vehicle-to-Everything} (V2X) creates many new opportunities in the area of wireless communications, while its feasibility on enabling \emph{vehicular positioning} (VP) has not been fully explored despite its importance for autonomous driving. This article aims at investigating whether V2X can help VP from different  perspectives. We first explain V2X's critical advantages over other technologies (e.g., GPS, RADAR, LIDAR, and camera) and suggest new scenarios of V2X-based VP. Then, we review  the state-of-the-art positioning techniques discussed in different standardizations and point out their limitations. Lastly, some promising research directions for V2X-based VP are presented, which contribute to realizing fully autonomous driving by overcoming the current obstacles.
\end{abstract}


\section{Introduction}\label{Introduction}

\emph{Autonomous driving} (auto-driving) has been a long desired technology for improving our productivity and safety by automating transportation and thereby freeing human drivers from their tasks. 
With rapid advancements in multidisciplinary technologies, state-of-the-art auto-driving technologies are being quickly transferred from labs to our real lives. It is expected that conditional auto-driving within limited areas (e.g., highway) will be available by $2022$, and fully auto-driving will be realized by $2025$.

One essential operation in auto-driving is positioning, namely recognizing the car's absolute and relative positions concerning other objects such as buildings, pedestrians, and other vehicles. Auto-driving places much more stringent requirements on positioning accuracy than other services because an error  can lead to fatal accidents, as exemplified by recent cases involving Tesla's  and Uber's test cars. Specifically, \emph{vehicular positioning} (VP) is a challenging task due to a wide range of requirements, including high accuracy and reliability, ultra-low latency, and cost-efficiency.  

\begin{itemize}
\item {\textbf{Accuracy}: High accuracies have been  indispensable requirements that every VP technology must guarantee. Most of the latest cars support several driving assistant schemes based on high-precision positioning technologies, e.g., GPS, camera, RADAR, and LIDAR, which can achieve centimeters-level accuracies in favorable environments. 
For example, a car navigates its chosen route on a map as GPS informs its current location. Besides, \emph{adaptive cruise control} (ACC), adjusting the vehicle speed automatically to maintain the safety distance from a vehicle ahead, can be implemented with a camera, RADAR, and LIDAR. 
On the other hand, complex and dynamic driving conditions limit the usage of these technologies mentioned in the sequel. This is one reason for which explains why the current auto-driving remains at the beginner level. }
\item \textbf{Reliability}: VP should always be available without an outage, but surrounding environments are continuously changed due to high mobility and may become unfavorable. For example, even though GPS is widely used for absolute positioning, it does not work in urban canyon-like environments where \emph{Line-of-Sight} (LoS) links to GPS satellites are often blocked. The relative positioning using cameras, RADAR, and LIDAR are valid only when LoS to targets are present. 

\item \textbf{Latency}: 
The current state-of-the-art technology, LIDAR, suffers from the long latency caused by collecting a large amount of data from scanning the surrounding environment and processing the data. Though it is  capable of achieving centimeter-level accuracy and widely used for mapping, its application to real-time positioning is problematic due to the scanning and processing latency. 

\item \textbf{Cost}: Automobile manufacturers are struggling with the high cost of auto-driving technologies.  In particular, many onboard sensors and powerful processing units are needed for accurate positioning, causing a high development cost. For example, according to an industry leader Velodyne, a LIDAR costs about $75,000$ USD. 
\end{itemize}

Recently, \emph{Vehicle-to-Everything} (V2X) has emerged as a new type of vehicular communications comprising \emph{Vehicle-to-Infrastructure} (V2I), \emph{Vehicle-to-Vehicle} (V2V), and \emph{Vehicle-to-Pedestrian} (V2P) communications \cite{seo2016lte}. Along with the recent evolution of wireless networks termed \emph{New Radio} (NR), V2X can provide many new services for auto-driving. One main direction is V2X-based positioning that has not been actively explored yet. In this article, we first begin with a discussion on how V2X can help VP. There exist a few recent works dealing with VP in NR, especially focusing on \emph{millimeter-wave} (mmWave) bands (see e.g., \cite{wymeersch20175g}). In this article, we aim at investigating VP from the V2X perspective covering all possible frequency bands of V2X. Then we introduce the latest standardization activities of positioning in different organizations and list open technical challenges. Finally, we point to several promising directions in which solutions for overcoming the current challenges can be found  and summarize relevant research issues and opportunities.

\section{V2X-Based Vehicular Positioning}
In this section, the key advantages of V2X are first discussed in the context of fulfilling the requirements mentioned above. 
Then, V2X's possible frequency bands are compared to investigate the feasibility of V2X-based VP.  
Last, the  scenarios where V2X-based VP finds its strengths are described.

\subsection{Advantages of V2X for Vehicular Positioning}
V2X has the following distinct features for overcoming the drawbacks of the conventional technologies mentioned above. 

\subsubsection{\textbf{Robustness of V2X channel}} Recall the reliability issue of GPS, RADAR, and LIDAR, where frequent outages occur in cases with LoS blockage. 
 On the other hand,  frequency bands of V2X channels, which are verified to be valid in these environments, can improve the reliability of VP. The in-depth discussion of the V2X channel's feasibility on VP is presented in Sec. \ref{Sec:FrequencyBands}.  
 
\subsubsection{\textbf{Communication-aided positioning}} An autonomous vehicle needs to estimate  not only the location but also other information of a target (e.g., speed, size, and shape), which places a heavy burden on its processing units when we use RADAR or LIDAR relying on the reflected signals from blind targets. Consequently, the resultant positioning accuracy decreases, and the latency increases. With V2X, on the other hand, time-invariant information like the target's shape and size can be delivered via reliable wireless communications. Then, the processing unit entirely focuses on estimating time-varying details, such as the vehicle's location, leading to estimating a more accurate location with shorter latency.
 
\subsubsection{\textbf{Infrastructure-reuse}} Most signal processing techniques of positioning are similar to those of communications, e.g., detection, filtering, and beamforming. It is thus possible to reuse the existing V2X infrastructure such as \emph{base stations} (BSs) and \emph{road side units} (RSUs) with software only upgrading, which is a cost-effective solution \cite{Peral-Rosado2019}. 

\begin{figure}
\centering 
{\includegraphics[width=7.0cm]{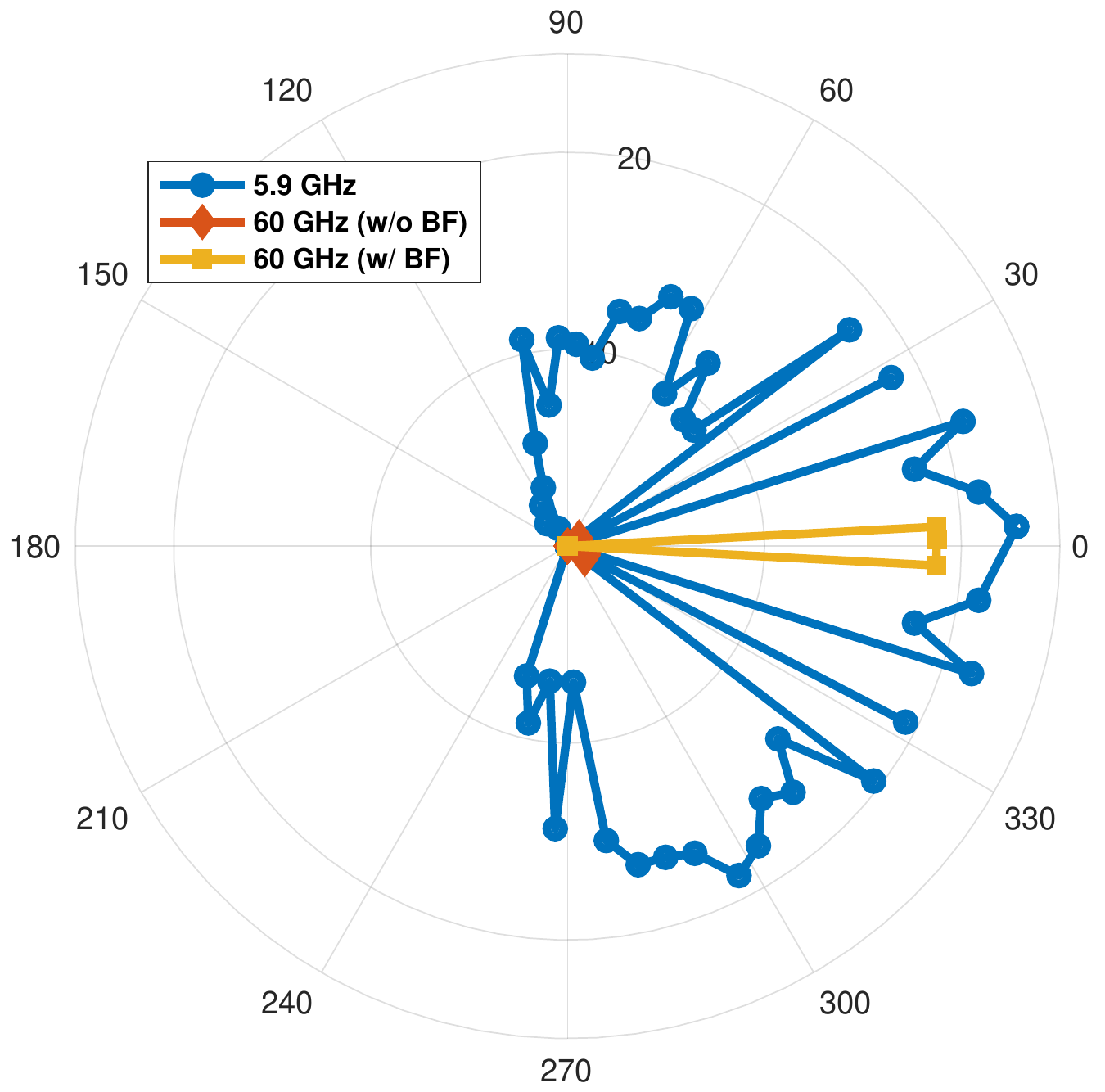}} 
\caption{V2X channels' RSS (in dB) in different arrival directions when the inter-vehicle distance is $100$ m. The V2V channel model in an urban LoS is considered \cite{3GPPTR37885}. The transmit power is $23$ dBm, noise power spectral density is -$174$ dBm/Hz, and system bandwidth is $100$ MHz. The number of receive antennas for beamforming is $64$. }
\label{RSSvsAoA}
\end{figure}

\begin{figure*}
\centering 
{\includegraphics[width=13cm]{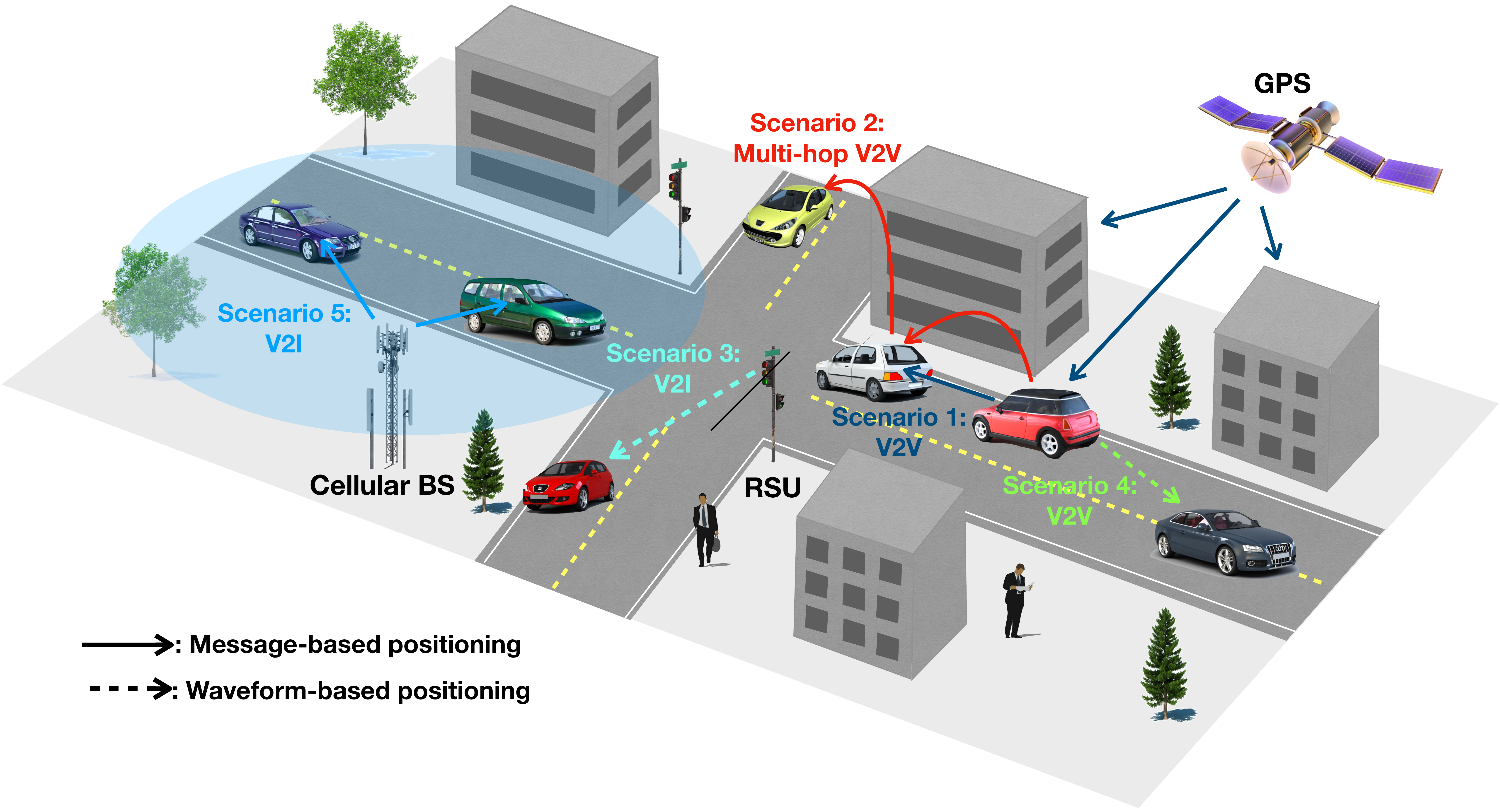}} 
\caption{Operation scenarios in V2X-based vehicular positioning}\label{Overview}
\end{figure*}

\subsection{Operating Frequency Bands}\label{Sec:FrequencyBands}
ITU-R recommends  two frequency bands for V2X, namely $5.9$ GHz and $60$ GHz bands, whose applicable VP services can be different depending on their propagation properties. 

Fig. \ref{RSSvsAoA} shows V2X channel's \emph{received signal strength} (RSS) in different arrival directions when the $5.9$ GHz and $60$ GHz bands are considered. The $60$ GHz band's signal attenuation with distance is more severe than the $5.9$ GHz signal. Multi-antenna beamforming is inevitable to compensate for the difference. However, most paths are likely to arrive within a short delay-spread (around $5$--$30$ $\mu$sec). Therefore, only a few paths' strengths are amplified by beamforming, making the paths visible but remaining ones invisible. This sparsity helps identify the visible paths precisely. 
As a result, the $60$ GHz band can achieve accurate positioning if dominant paths exist, e.g., LoS environments.

For the $5.9$ GHz band, on the other hand,  RSS is sufficiently large without beamforming, and many signal paths can be observed in different directions. These paths include NLoS paths via scattering, diffracting, and \emph{et al}.  These are almost always found regardless of the existence of an LoS path, which can become clues for NLoS positioning. In the area of positioning, the nature of NLoS multi-paths traditionally has been regarded as a negative factor to hamper accurate positioning, and most works focus on NLoS path identification to discard them. On the other hand, a few recent studies on NLoS positioning (see, e.g., \cite{miao2007positioning}) aims at exploiting NLoS signal paths can enhance positioning performance. 

In summary, V2X  can provide seamless VP services in both LoS and NLoS environments by exploiting different features of the $5.9$ GHz and $60$ GHz bands.

\subsection{Scenarios of V2X-Based Vehicular Positioning}

Depending on whether vehicles' locations are communicated or estimated, we present two types of  positioning approaches. First, \emph{message-based positioning} is to let the car know the position of itself or its neighbors by receiving the message containing the explicit location information via  V2X communications. Second, \emph{waveform-based positioning} is to estimate the locations  based on the physical relations among several measurement results, termed positioning elements in this article. The positioning elements can be estimated by detecting a waveform referring to a specific signal shape already known by the vehicle. The particular positioning elements and techniques are elaborated in Section \ref{Sec: NRPositioning}.

Based on the two, we introduce the operating scenarios of V2X-based VP, as illustrated in~Fig.~\ref{Overview}.

\vskip 0pt
\noindent \textbf{Scenario 1}. The absolute or relative positions detected by onboard GPS, RADAR, or LIDAR \emph{et al.} can be transmitted to neighbor vehicles via direct V2V transmissions.  

\noindent \textbf{Scenario 2}. The absolute or relative positions detected by onboard GPS, RADAR, or LIDAR \emph{et al.} can be transmitted to vehicles in NLoS via multi-hop V2V transmissions.
  
\noindent \textbf{Scenario 3}. A vehicle's absolute position can be estimated by detecting waveforms transmitted by anchors, which are the infrastructure whose locations are known in advance, i.e., BSs or RSUs.

\noindent \textbf{Scenario 4}. The relative position of a vehicle can be estimated by detecting the waveforms  transmitted by other vehicles.

\noindent \textbf{Scenario 5}. A macro BS broadcasts the entire positioning map within its coverage, which can be made through measurement reports of vehicles obtained in Scenarios 1-4. 
\vskip 0pt

Message-based positioning is used in Scenarios 1, 2, and 5, while  waveform-based positioning is used in Scenarios 3 and 4. In the past decade, the former has been considered as a standard approach for VP, creating an active research area named cooperative positioning \cite{alam2013cooperative}. 
On the other hand, the latter is a relatively new direction in the area of VP, while it has been studied in indoor and cellular localizations. Thus, we mainly discuss the waveform-based~VP. 

\section{State-of-the-Art of Positioning Techniques}\label{Sec: NRPositioning}
This section reviews the state-of-the-art positioning techniques in different standardization groups. First, we explain key positioning elements as a preliminary. Then, we introduce the recent standardizations of positioning techniques and describe its limitations when applied to VP. 

\subsection{Introduction to Key Positioning Elements}
Depending on means used to capture a specific physical property of radio signals, positioning elements are categorized as follows. 

\subsubsection{\textbf{Time-based}}  Time-based elements use one basic theory of classical physics that a signal propagation speed is constant at light speed $c=3\cdot 10^8$  (m/sec). \emph{Time-of-Arrival} (ToA) refers to the flight-time of a radio signal between an anchor and a \emph{User Equipment} (UE), which can be translated into the corresponding flight-distance by multiplying $c$. The relation relies on the assumption of anchor-UE synchronization, which is sometimes infeasible in practice.   To cope with it,  \emph{Time-Difference-of-Arrival} (TDoA) is introduced, referring to the time difference between two radio signals' arrivals originated from or received at different anchors. In the same vein, TDoA is equivalent to the difference of the flight-distance from the vehicle to the two anchors. Geographically, a ToA represents a circle for the UE's possible position due to a radio signal's isotropic propagation. On the other hand, a TDoA represents a hyperbolic curve because it represents the difference between two Euclidean distances. 

\subsubsection{\textbf{Phase-based}} A phase is another factor that regularly varies  with the corresponding frequency. \emph{Phase-of-Arrival} (PoA) is the phase change from the initial value, which is proportional to the corresponding flight-distance. Due to the nature of periodicity, the available range without ambiguity is the reciprocal of the concerned frequency, which is too short of use in the practical frequency range of V2X. It is thus recommended of using \emph{Phase-Difference-of-Arrival} (PDoA) defined as the phase difference between two arrival signals with different frequencies. The available range of PDoA is the reciprocal of the frequency gap, which is significantly larger than the former. 
\subsubsection{\textbf{Angle-based}} Due to  the advance of signal processing techniques and the appearance of massive antenna arrays, it is possible to find out a radio signal's propagation directions precisely, namely  \emph{Angle-of-Arrival} (AoA) and  \emph{Angle-of-Departure} (AoD). For instance, consider a signal departing from a single transmit antenna. Based on the far-field assumption, the received signals' phase gaps at different receive antennas depend on AoA. Conversely, signals originating from different transmit antennas cause a phase gap of the received signal, according to AoD.

\begin{table*}
\centering
\caption{Comparison of State-of-the-Art Positioning in Different Standards}
\setlength{\tabcolsep}{2pt}
\footnotesize
\begin{tabular}{c|c|c|c|c|c|c|c|c}
\toprule
\multicolumn{2}{c|}{} & \multicolumn{3}{c|}{\textbf{3GPP}} & \multicolumn{2}{c|}{\textbf{IEEE}} & \textbf{ETSI} & \textbf{SAE} \\ \midrule
\multicolumn{2}{c|}{\textbf{Specifications}} & \multicolumn{3}{c|}{{Rel.~16}} & \multicolumn{2}{c|}{802.11bd} & {TS 103.301} & {J2945/1} \\ \midrule
\multicolumn{2}{c|}{\textbf{Key Techniques}} & ECID & OTDoA & UTDoA & DSRC & FTM & GPC & RSU-Based \\ \midrule
\multicolumn{1}{c|}{\multirow{2}{*}{\textbf{Approach}}} & \textbf{Message-Based} &  & &  & $\checkmark$ &  & $\checkmark$ &  \\
\multicolumn{1}{c|}{} & \textbf{Waveform-Based} & $\checkmark$  & $\checkmark$ & $\checkmark$ & & $\checkmark$ &  & $\checkmark$ \\ \midrule
\multirow{2}{*}{\textbf{Positioning Elements}} & \textbf{Time-Based} & $\checkmark$ & $\checkmark$ & $\checkmark$ &  & $\checkmark$ &  & $\checkmark$ \\ 
 & \textbf{Angle-Based} & $\checkmark$  &  &  & &  &  &  \\ \midrule
\multirow{5}{*}{\textbf{Limitations}} & \textbf{Relying on GPS} &  &  & & $\checkmark$ &  & $\checkmark$ &  \\ 
 & \textbf{Network Sync.} & & $\checkmark$  & $\checkmark$ &  &  &  \\
 & \textbf{Disable in NLoS} & $\checkmark$ & $\checkmark$ & $\checkmark$ & &  $\checkmark$ &  &  \\  
 & \textbf{Prone to High Mobility} & $\checkmark$ & $\checkmark$ & $\checkmark$ & $\checkmark$ & $\checkmark$ & $\checkmark$  & $\checkmark$\\  
\bottomrule
\end{tabular}\label{Comparison_Standardization}
\end{table*}

\subsection{Standardization}\label{StandardVP}
There are ongoing positioning works in different standard organizations such as \emph{The 3rd Generation Partnership Project} (3GPP), \emph{Institute of Electrical and Electronics Engineers} (IEEE), \emph{The European Telecommunications Standards Institute} (ETSI), and \emph{The Society of Automotive Engineers} (SAE). 
We compare their key features in Table \ref{Comparison_Standardization} and describe them as follows.

\subsubsection{\textbf{3GPP}} 

3GPP NR is expected to combine multiple dimensions with improving positioning accuracy substantially 
by exploiting a wide range of operating frequencies and utilizing massive antenna arrays. Recently, a new \emph{work item} (WI) on NR positioning support was initiated in 3GPP Rel.~16. 
This WI's priority is to evaluate and specify the feasibility and scalability of the existing RAT-dependent positioning techniques, including \emph{Enhanced Cell-Identifications} (ECID), \emph{Observed-TDoA} (OTDoA), and \emph{Uplink-TDoA} (UTDoA). 

ECID is a Cell-ID based method where the UE's position corresponds to the geographical coordinates of the serving BS obtained by using tracking-area-update or paging. To improve the accuracy, ECID uses \emph{Round-Trip Time} (RTT) between the UE and the serving BS, which give a circle for the UE's possible location. A unique positioning is possible if AoA can be measured. 

OTDoA is based on the downlink TDoA measurements of \emph{Positioning Reference Signals} (PRSs) from multiple BSs. The measured TDoAs are fed back to the location server via the serving BS for calculating the UE's location. Recall that one TDoA from two BSs gives a hyperbolic curve whose two foci are the corresponding BSs' locations. At least three BSs are required for obtaining a unique position to find the intersection of the hyperbolic curves. 

UTDoA is an uplink counterpart to OTDoA standardized in 3GPP Rel.~11. To enable the TDoA measurement at  BSs, a \emph{Sounding Reference Signal} (SRS) is used, which is an uplink reference signal  for uplink channel quality estimation. Since a UE's transmit power is limited, its performance is worse than OTDoA. On the other hand, BSs measure TDoA instead of UE, and the computation complexity impact on the UE is less than OTDoA. It helps UE's energy savings and lifetime extension. 

Unfortunately, the requirements of VP are much more demanding than those of the WI. For example, in the WI's commercial use cases, horizontal and vertical positioning errors for outdoor UEs are respectively  $10$ m and $3$ m. However, a much higher level of accuracy is needed for VP, such as $0.1$ m lateral error and $0.5$ m longitudinal error for platooning \cite{TR22.886}. 
We explain several technical limitations as follows. 

\begin{itemize}
\item A common drawback of the above positioning techniques is the low resolution due to limited bandwidth. In principle, the bandwidth is proportional to the time resolution. Recently, FCC decided to allocate $20$ MHz of  $5.9$ GHz band for V2X technologies \cite{FCC_Decision}; a waveform with $20$ MHz bandwidth yields $50$ nsec resolution, which is translated to $15$ m positioning error when considering the speed of light $c$. Exploiting the broad bandwidth of a mmWave band leads to increasing time-resolution, but its allocation discussion for V2X has not yet begun. 
\item OTDoA and UTDoA rely on the critical assumption that all anchors are perfectly synchronized. However, there inevitably exist  synchronization errors among anchors due to some practical reasons, e.g., signaling overhead, finite fiber links, and clock misalignment. Besides, due to network densification, more small-size BSs such as picocells and femtocells are likely to be deployed in a plug-and-play manner, making the synchronization more difficult. The WI  considers the maximum synchronization gap between BSs as $50$ nsec \cite{SID-NR-Positioning}, corresponding to a $15$ m error that is unacceptable for auto-driving.
\item The conditions for these techniques to have a unique positioning is that the number of anchors with a LoS link should be at least the minimum (i.e., $3$ for OTDoA and UTDoA). Unfortunately, the number is random, depending on the  current surrounding environment, and can be less than the minimum. It is vital to develop a VP technique working in an NLoS environment called \emph{hidden vehicle positioning}. 
\item These techniques require several transmissions between anchors and a UE, which is vulnerable to high mobility. In the case of OTDoA, vehicles keep moving during multiple PRS receptions, and there may exist misalignment between real and estimated positions. Besides, a vehicle's fast velocity results in a Doppler shift, making the problem more challenging.      
\end{itemize}

\subsubsection{\textbf{IEEE}} 
Though \emph{Dedicated Short Range Communication} (DSRC) has been widely used in many countries as a standard of V2X communications, its limit has been reached as vehicular applications  become complex, and the requirements are demanding. The performance of DSRC is acceptable for basic safety applications, of which the end-to-end latency requirement is around $100$ msec. On the other hand, the latency requirements of advanced vehicular applications are much more stringent, i.e., $3$-$100$ msec for advanced driving and $5$ msec for auto-driving \cite{TR22.886}. 

Recently, a new task group has been formed  to develop a new standard, IEEE 802.11bd, as an evolution of DSRC. VP is one main objective of 802.11bd, and it is likely to use the positioning scheme in 802.11az, another ongoing standard known as \emph{Next Generation Positioning} (NGP).  Its core  feature is \emph{Fine Time Measurement} (FTM), which was firstly introduced in 802.11mc. FTM is RTT-based ranging protocol. After a few handshake procedures, it provides a high-resolution delay estimation at picosecond granularity. The goal of 802.11az is to propose some improvements in FTM, such as multi-user ranging, trigger-based ranging, AoA/AoD measurements, and \emph{et al}. On the other hand, handshake signaling causes a certain level of delay. According to \cite{Banin2017}, it requires $30$ ms to measure one RTT, and $100$-$120$ ms delay occurs to obtain three RTTs for positioning. When a vehicle is moving with a speed of $100$ km/h, a typical error is $2.8$-$3.4$~m. Its current target application is thus limited to mobile localization, and application to VP is still under discussion. 
 
\subsubsection{\textbf{ETSI}}
VP in ETSI is based on \emph{cooperative awareness},  where vehicles share their absolute positions obtained from GPS by exchanging \emph{cooperative awareness messages} (CAMs) periodically  \cite{ETSI103301}. However, the poor GPS accuracy in practical driving scenarios is insufficient for VP. Besides, discontinuous CAM transmissions cause position misalignment between two successive CAM receptions called \emph{inter-reception time}~(IRT), i.e., $2$ m error for $10$ Hz CAM when a vehicle is moving at $20$ m/s speed. Several approaches have been suggested to overcome these limitations.  First, \emph{GNSS positioning correction} (GPC) is considered, where RSUs or BSs generate GPS correction data for vehicles. Second, it is possible to reduce such error according to the fusion of different positioning-related information, i.e., neighbor's absolute and relative position, and onboard GPS estimates. Third, a lightweight awareness message is designed to reduce the IRT. 

\subsubsection{\textbf{SAE}} SAE aims at addressing the challenges of unreliable GPS estimates without an open sky. Especially, SAE focuses on investigating the applicability of RSU-assisted enhanced positioning based on the existing DSRC and FTM procedure. It includes several operating scenarios, such as trilateration using $3$ RSUs and augmentation using one RSU and two GPS satellites. However, it has similar issues with IEEE since they share the same radio interface. 

\begin{figure}
\centering 
{\includegraphics[width=8cm]{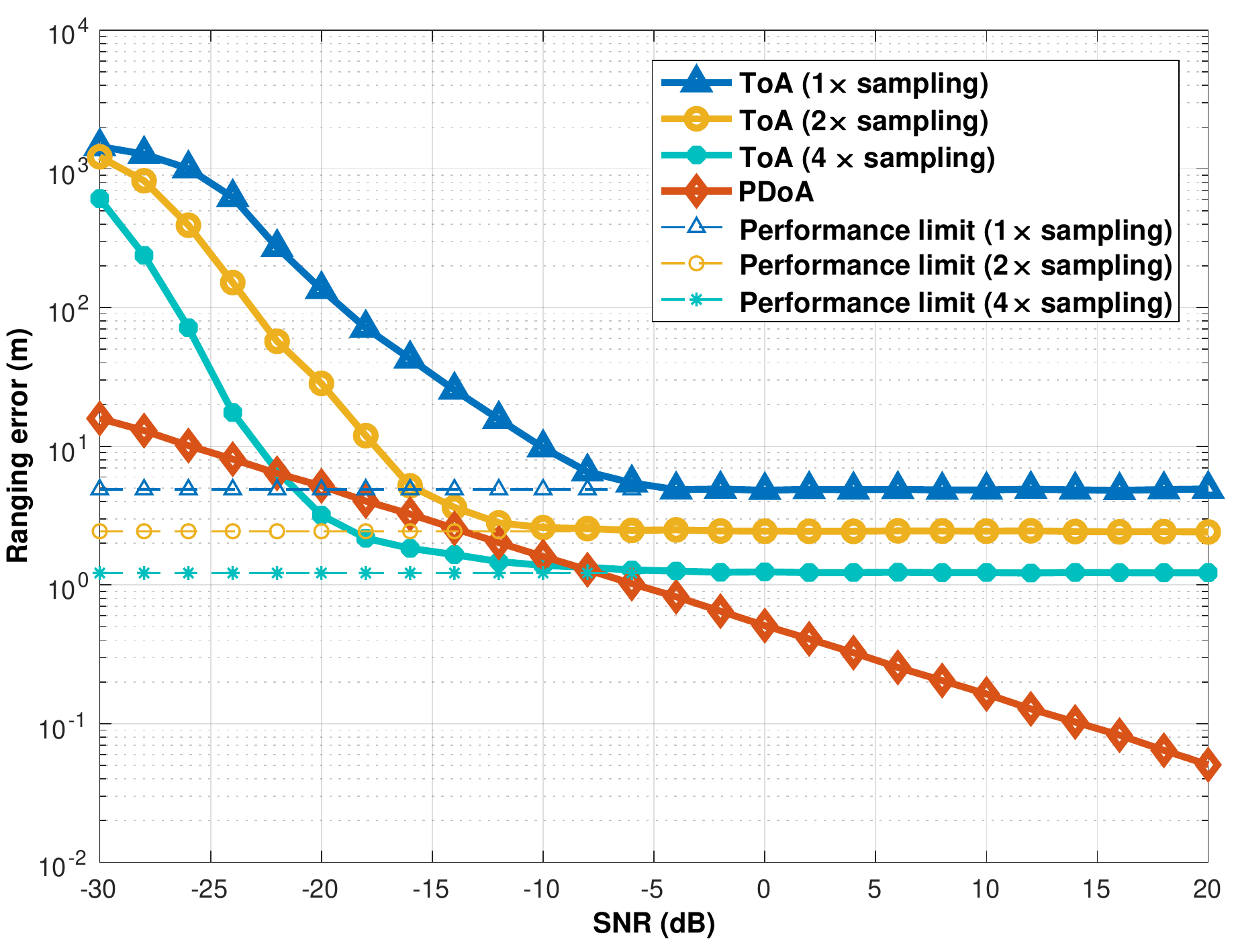}} 
\caption{ToA and PDoA's ranging errors under an AWGN channel. We consider a NR frame structure based on \emph{Orthogonal Frequency Division Multiplexing} (OFDM).  $2048$ subcarriers are used with $15$ KHz frequency spacing. We use the base sampling rate ($15$KHz $\times$ $2048$ = $30.72$ MHz) and two oversampling rates ($61.44$ MHz and $122.88$ MHz).  Assuming that a sampling error is uniformly distributed within $[0,  \frac{c}{\textrm{sampling rate}}]$ where $c$ is the light speed, the performance limit of ToA is calculated as $\frac{1}{2} \frac{c}{\textrm{sampling rate}}$. For PDoA, we use two tones in $1$st and $12$th subcarriers. 
}
\label{PDoA}
\end{figure}

\section{Research Directions for V2X-Based Vehicular Positioning}

This section aims at giving several promising directions of V2X-based VP, which deserve consideration for overcoming the limitations of state-of-the-art. 

\subsection{Using Phase-Based Elements for Band-limited Scenarios}

All positioning techniques mentioned above rely on time-based elements (i.e., ToA and TDoA), which are discontinuous information determined by the concerned sampling interval. The resultant positioning accuracy is thus limited if the allowable bandwidth is narrow (i.e., $20$ MHz in $5.9$ GHz \cite{FCC_Decision}). On the other hand, phase-based elements, which are continuous information over $2\pi$, can embed more information than time-based ones. They can provide acceptable positioning accuracy in band-limited situations. For example, PDoA, the product of the propagation delay and the difference between the two frequencies, is valid unless the two frequency tones are the same. Noting that the frequency gap can be regarded as the  bandwidth,  PDoA-based positioning is likely to be more resource-efficient than time-based ones. 
As a result, it can be beneficial in terms of radio resource management and collision avoidance since the radio resources required for a single vehicle can be reduced.

Fig.~\ref{PDoA} shows the performance comparison between ToA- and PDoA-based ranging under an AWGN channel. It is shown that ToA-based ranging always has a performance limit in the high SNR regime. Two types of ranging errors exist. The first is the estimation error due to noise, which diminishes as SNR increases. The second is the sampling error due to the limited bandwidth, which remains constant regardless of SNR.  
Oversampling is one common approach to overcome the sampling error, as shown in Fig.~\ref{PDoA}. However, it requires a faster \emph{analog-to-digital} (ADC) converter, resulting in high manufacturing cost. Besides, a V2X architecture is mainly designed for communication, but the oversampling has no advantage in communication performance. As a result, oversampling of ToA ranging is impractical for V2X-based VP.  For PDoA-based one, on the other hand, the ranging error keeps decreasing as \emph{Signal-to-Noise Ratio} (SNR) increases. The oversampling is not required, and it can be implemented cost-effectively without modifying a communication-based architecture. 
For the approach to be practical, some issues are summarized~below.
\subsubsection{\textbf{Phase ambiguity}} Recall the ambiguity issue where PDoA becomes the same in every interval of  $\frac{c}{\Delta}$ where $c$ is the speed of light,  and $\Delta$ is the frequency gap. The maximum distance estimation range is thus limited to avoid the ambiguity. It is overcome by a hierarchical scheme based on multi-frequencies. First, the distance to a target is roughly estimated by using two adjacent frequencies, which makes it possible to reduce the estimation range. Next, it is fine-tuned by using two frequencies with a more significant difference within the reduced range. 
\subsubsection{\textbf{Frequency-selective channel}} As the signal bandwidth increases, more channel taps are observed, and the channel becomes frequency-selective. It hampers the accurate detection of PDoA since these multi-path signals with different propagation paths are non-coherently combined, and the resultant phase is distorted. To address it, we can use channel estimation information. The multi-path delay profile of the concerned channel helps the cancellation of phase components contributed by selective fading.

\subsection{A Multi-Path  Channel Helps Hidden Vehicle Positioning}

\begin{figure}
\centering 
{\includegraphics[width=8cm]{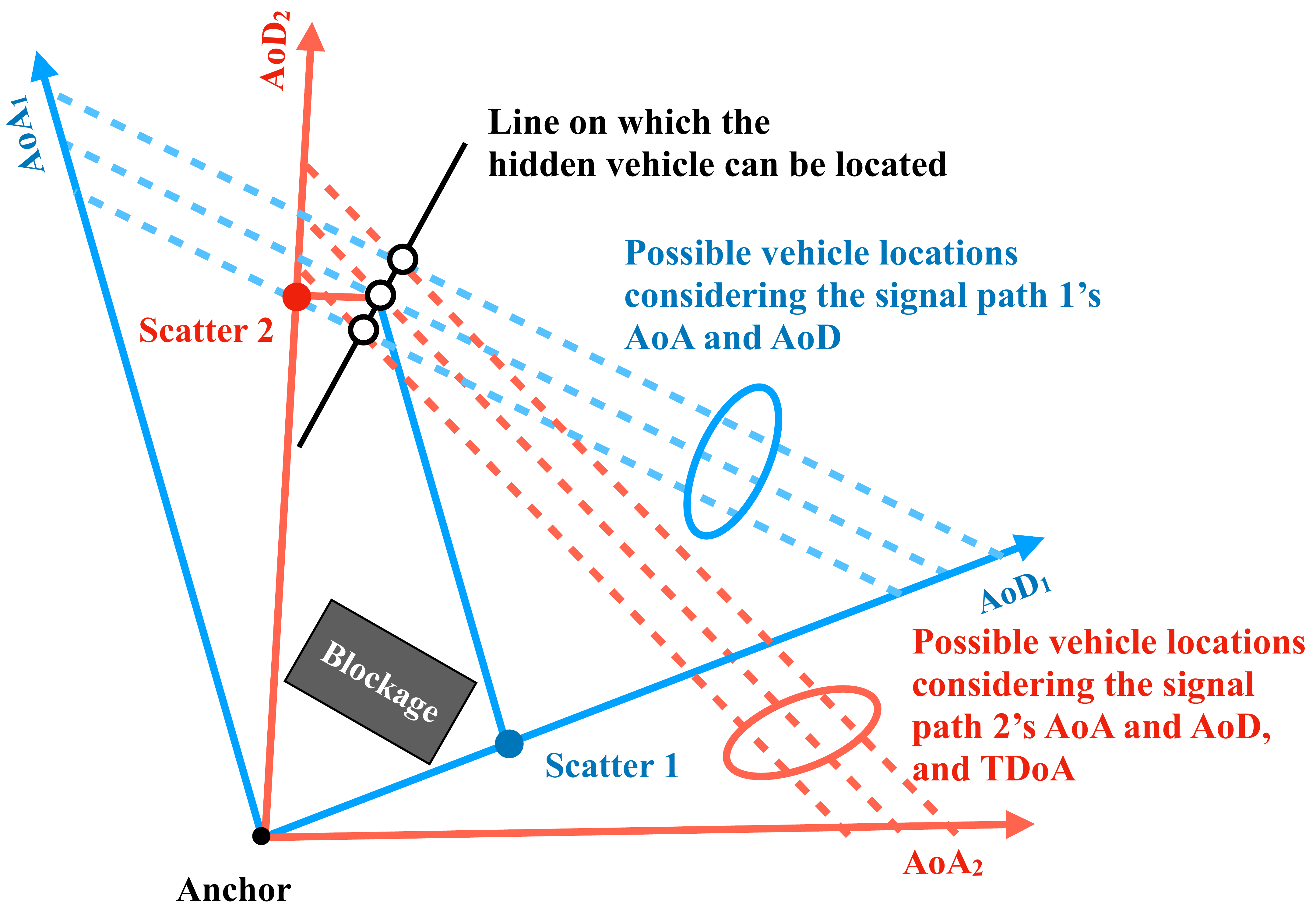}} 
\caption{An illustration of hidden vehicle positioning when two signal paths are detected with AoAs, AoDs, and TDoA. 
}
\label{HiddenVehiclePositioning}
\end{figure}

The multi-path nature of a wireless channel can open a new dimension for VP \cite{han2018sensing}. When an anchor broadcasts a waveform, its multiple replicas can be delivered to a hidden vehicle  through different signal paths. It enables the car to estimate an individual signal path's positioning elements. By the interplay of these positioning elements, we can find the hidden vehicle's location. 

Consider a 2D single-bounce multi-path channel model, where each path has one scatter. All signals depart from a single origin (anchor) and arrive at a destination (hidden vehicle), making it possible to form a bilateral relation between them through the multiple positioning elements. We consider  that two signal paths are detected with AoDs, AoAs, and TDoA, as in Fig. \ref{HiddenVehiclePositioning}. 

Assume that path $1$'s flight distance is given. Along with its AoA and AoD, we can make a line representing the vehicle's possible location (dotted blue line), whose length is equivalent to the flight distance. Second, path $2$'s flight distance can be calculated by adding the TDoA to path $1$'s flight distance, and we can make another line (dotted red line) concerning path $2$. The pair of the lines have one crossing point.  Changing path $1$'s propagation distance makes the trajectory of the crossing point (solid black line). Adding one more signal path generates one more such a line, we can infer the vehicle's location by finding the intersection between the two. 

The multi-path-geometry approach has many potential advantages to overcome the limitations mentioned in Sec. \ref{StandardVP}.  First, by combining multi-path signals, it is possible to cancel out the error in each signal estimation, yielding more accurate positioning than a single-path approach.  Second, all signals are perfectly synchronized because they depart from one anchor in the same instant. Third, it is performed by one-way transmission without a feedback channel, leading to reducing  the positioning duration. The resultant error due to high mobility can be minimized. 

On the other hand, several issues should be addressed for practical use. 

\begin{figure}
\centering 
{\includegraphics[width=8.7cm]{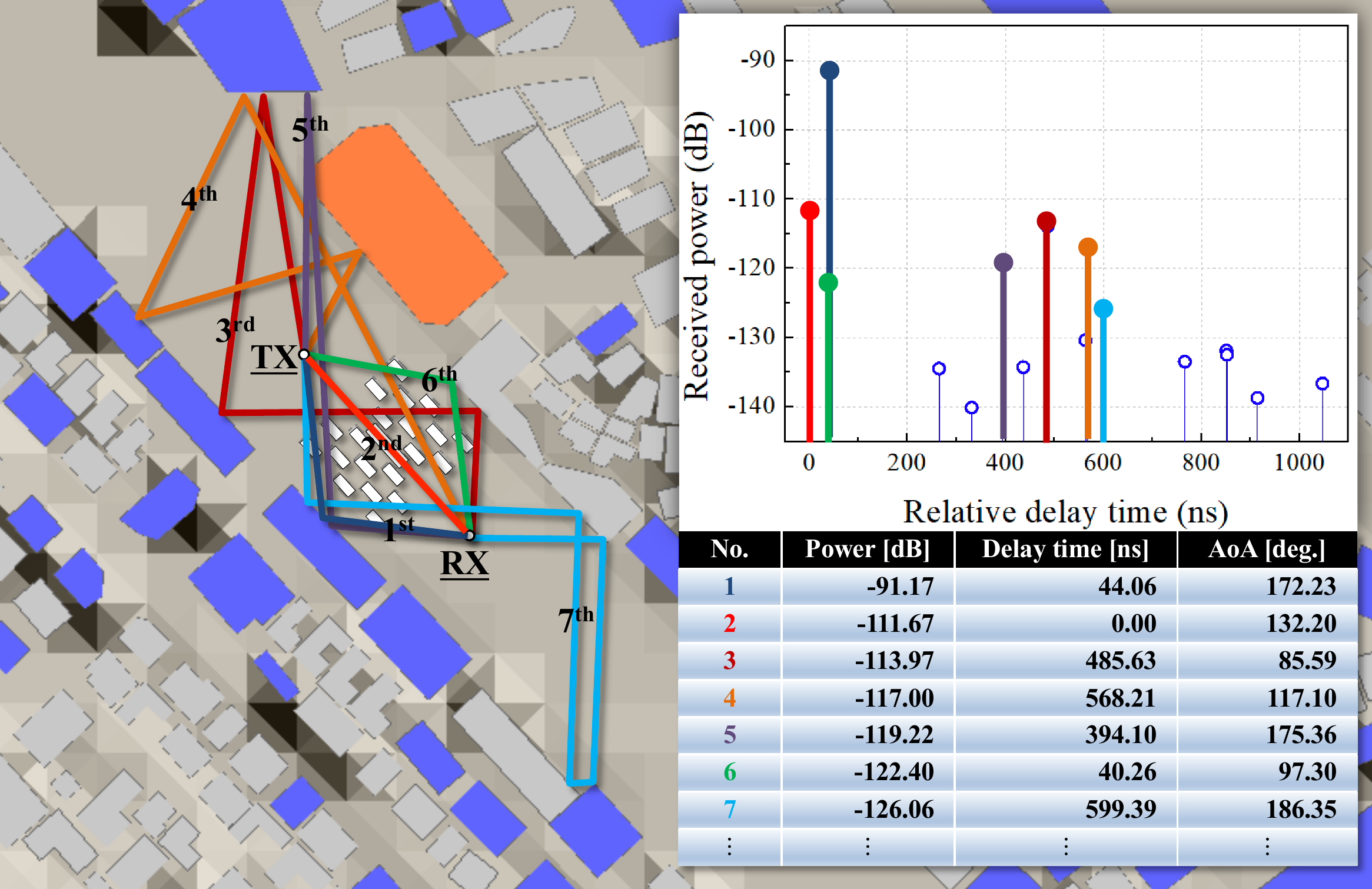}} 
\caption{A ray-tracing result of a $6$ GHz frequency band in a real NLoS environment (Seodaemun-Gu, Seoul, Korea).  Different colors are used to display signal paths with the top $7$ received power. The first one represents a path diffracted over the blockage's roof, and the $2$nd and $6$th paths are one-bounce ones. The others are multi-bounced paths due to several reflections. Readers interested in our ray-tracing simulations can check http://tiny.cc/x176jz.
}
\label{Raytracing}
\end{figure}

\subsubsection{\textbf{Coexistence of multi-bounce paths}} 
Fig. \ref{Raytracing} represents a ray-tracing simulation result in a real NLoS environment. It is observed that multi-bounce paths due to reflections and diffractions coexist, making it much more complex to find a hidden vehicle's location by concerning higher-order geometries. Besides, the received power is unreliable to identify a single-bounce path (See $6$th path).  
Two directions are suggested to address it. First, single-bounce paths can be identified by utilizing an AoA-AoD relation that they point to a common scatter if it is a single-bounce path. Map information helps determine if there is a scatter in that position \cite{Kanhere2019}. Second, a new classification algorithm can be developed, where a centroid of many candidate points made by different path combinations becomes the vehicle's estimated location.  
\subsubsection{\textbf{An insufficient number of signal paths}} The feasible condition of this approach is that a certain number of signal paths should be detected (e.g., $3$ paths in Fig. \ref{HiddenVehiclePositioning}). 
As observed in Fig. \ref{Raytracing}, however, the number of observable single-bounce paths is random and can be less than $3$. A way to overcome this issue is to combine signal paths observed at different times until the sufficient number of signal paths is collected.  Note that it requires the vehicle's maneuver information. 
To this end, it is required to incorporate vehicles' movement tracking and prediction within VP framework.  

\subsection{Using Backscatter Tags as Cost-Effective Anchors}

Anchor densification can reduce the NLoS situations while paying for the high deployment cost. Instead of typical anchors (e.g., BS or RSU), one economical solution is to deploy backscatter tags such that a reader-mounted vehicle senses nearby tags. Then, the backscatter tags feed back their IDs, corresponding to their locations, to the reader by modulating and reflecting the incident waveform. A key feature of backscatter technology is to wirelessly power many positioning tags, relieving the burden of battery charging. Besides, it helps reduce the deployment and production costs due to its small form factor and simple architecture without energy-hungry components.

We can deploy backscatter tags on a road, enabling a vehicle to know its own location by reading the tag's ID when passing over it (e.g.,~\cite{Qin2017}). However, this ID-based design has a disadvantage that as heavy vehicles pass over the tags, they easily break down, and the maintenance cost increases. To overcome the drawback, a new deployment plan is proposed in which backscatter tags are installed alongside the road \cite{Backscatter_pos_2019}. It makes the maintenance of tags much easier by prolonging their lifetimes. Some directions are suggested to make it  practical.

\subsubsection{\textbf{Location mismatch between tags and a vehicle}} The proposed tag deployment does not guarantee that the tag's position is equivalent to the vehicle's one. To compensate for the difference, not only the tag's ID but also the relative position should be obtained. In other words, the joint communication-and-sensing design is essential to get both of them simultaneously. 
\subsubsection{\textbf{New multi-antenna beamforming}} Due to the double-propagation of a backscatter channel, its attenuation is much more severe than other transmission technologies. Multi-antenna beamforming can overcome this limitation by forming a sharp beam in a specific direction. On the other hand, it decreases the reader's coverage, and the resultant contact time becomes short. Therefore, a new beamforming technology should be developed to optimize the tradeoff. 

\section{Conclusion}
In this article, we have investigated V2X as a pivotal technology and suggested several attractive directions to meet the VP's stringent safety requirements. This article can provide some useful instructions to realize fully auto-driving by  overcoming current obstacles. Besides, We can leverage the key features of V2X from a positioning perspective to open up new research areas such as \emph{unmanned aerial vehicle} (UAV) positioning and control. 

\bibliographystyle{ieeetr}
\bibliography{VehPos_Mag}

\begin{thebibliography}{10}

\bibitem{seo2016lte}
H.~Seo, K.~Lee, S.~Yasukawa, Y.~Peng, and P.~Sartori, ``{LTE} evolution for
  vehicle-to-everything services,'' {\em IEEE Comm. Mag.}, vol.~54, pp.~22--28,
  Jun. 2016.

\bibitem{wymeersch20175g}
H.~Wymeersch, G.~Seco-Granados, G.~Destino, D.~Dardari, and F.~Tufvesson,
  ``5{G} mm{W}ave positioning for vehicular networks,'' {\em IEEE Wireless
  Comm.}, vol.~24, pp.~80--86, Dec. 2017.

\bibitem{Peral-Rosado2019}
J.~A.~d. Peral-Rosado, G.~Seco-Granados, S.~Kim, and J.~A. Lopez-Salcedo,
  ``Network design for accurate vehicle localization,'' {\em IEEE Trans. Veh.
  Tech.}, vol.~68, pp.~4316--4327, May 2019.

\bibitem{3GPPTR37885}
{3GPP, TR 37.885, v15.3.0}, ``Study on evaluation methodology of new
  vehicle-to-everything (v2x) use cases for lte and nr,'' {\em Available:
  http://www.3gpp.org/DynaReport/37885.htm}.

\bibitem{miao2007positioning}
H.~Miao, K.~Yu, and M.~Juntti, ``Positioning for {NLOS} propagation: Algorithm
  derivations and {C}ramer-{R}ao bounds,'' {\em IEEE Trans. Veh. Tech.},
  vol.~56, pp.~2568--2580, Sep. 2007.

\bibitem{alam2013cooperative}
N.~Alam and A.~Dempster, ``Cooperative positioning for vehicular networks:
  Facts and future,'' {\em IEEE Trans. Intell. Transp. Syst.}, vol.~14,
  pp.~1708--1717, Dec. 2013.

\bibitem{TR22.886}
{3GPP, TR 22.886 v16.1.1}, ``Study on enhancement of {3GPP} support for {5G}
  {V2X} services,'' {\em Available: http://www.3gpp.org/DynaReport/22886.htm}.

\bibitem{FCC_Decision}
B.~Fletcher, ``{FCC} drives 5.9 {GHz} proposal for {C-V2X}, {Wi-Fi} use
  forward,'' {\em FierceWireless}, 12th Dec. 2019.

\bibitem{SID-NR-Positioning}
{3GPP, RP-182155}, ``Revised {SID}: Study on {NR} positioning support,'' {\em
  3GPP TSG RAN Meeting 81}, Sep. 2018.

\bibitem{Banin2017}
L.~Banin, O.~Bar-Shalom, N.~Dvorecki, and Y.~Amizur, ``High-accuracy indoor
  geolocation using collaborative time of arrival-whitepaper,'' {\em IEEE
  802.11-17/1397R0}, Sep. 2017.

\bibitem{ETSI103301}
{ETSI, TS 103.301, V1.3.1}, ``Intelligent transport system ({ITS}); vehicular
  communications; basic set of applications; facilities layer protocols and
  communication requirements for infrastructure services,'' Feb. 2020.

\bibitem{han2018sensing}
K.~Han, S.-W. Ko, H.~Chae, B.-H. Kim, and K.~Huang, ``Hidden vehicle sensing
  via asynchronous {V2V} transmission: {A} multi-path-geometry approach,'' {\em
  IEEE Access}, vol.~7, pp.~pp.169399--169419, Dec. 2019.

\bibitem{Kanhere2019}
O.~Kanhere, S.~Ju, Y.~Xing, and T.~S. Rappaport, ``Map-assisted millimeter wave
  localization for accurate position location,'' in {\em Proc. of IEEE
  GLOBECOM}, Dec. 2019.

\bibitem{Qin2017}
H.~Qin, Y.~Peng, and W.~Zhang, ``Vehicles on {RFID}: Error-cognitive vehicle
  localization in {GPS}-less environments,'' {\em IEEE Trans. Veh. Tech.},
  vol.~66, pp.~9943--9957, Nov. 2017.

\bibitem{Backscatter_pos_2019}
K.~Han, S.-W. Ko, S.~Lee, W.-S. Ko, and K.~Huang, ``Joint frequency-and-phase
  modulation for backscatter-tag assisted vehicular positioning,'' in {\em Proc
  IEEE SPAWC}, Jul. 2019.

\end{thebibliography}

\end{document}